\documentclass[aps,pra,superscriptaddress,twocolumn,amsmath,amssymb,floatfix]{revtex4}
\usepackage{graphicx}
\usepackage{dcolumn}
\usepackage{bm}
\usepackage{subfig}
\usepackage{caption}
\usepackage{amsmath}
\usepackage{amssymb}
\usepackage{array}
\usepackage{ulem}
\usepackage{float}
\begin{document}
\title{ Ultralong-Range Rydberg Molecules in a Divalent-Atomic System}
\author{B. J. DeSalvo}
\author{J. A. Aman}
\author{F. B. Dunning}
\author{T. C. Killian}
\affiliation{Rice University, Department of Physics and Astronomy, Houston, Texas, USA 77251}
\author{H. R. Sadeghpour}
\affiliation{ITAMP, Harvard-Smithsonian Center for Astrophysics, 60 Garden Street, Cambridge, Massachusetts 02138, USA}
\author{S. Yoshida}
\author{J. Burgd\"orfer}
\affiliation{Institute for Theoretical Physics, Vienna University of Technology, Vienna, Austria, EU}

\date{\today}

\begin{abstract}

We report the creation of ultralong-range Sr$_2$ molecules comprising one ground-state $5s^2$\,$^1S_0$ atom and one atom in a $5sns$\,$^3S_1$ Rydberg state for $n$ ranging from 29 to 36. Molecules are created in a trapped ultracold atomic gas using two-photon excitation near resonant with the $5s5p$\,$^3P_1$ intermediate state, and their formation is detected through ground-state atom loss from the trap. The observed molecular binding energies are fit with the aid of first-order perturbation theory that utilizes a Fermi pseudopotential with effective $s$-wave and $p$-wave scattering lengths to describe the interaction between an excited Rydberg electron and a ground-state Sr atom.

\end{abstract}

\maketitle

Low-energy scattering of the nearly free, excited electron in a Rydberg atom from a ground-state atom can bind the two atoms into an ultralong-range Rydberg molecule \cite{gds00,bbn09}. The resulting internuclear spacing is on the order of the size of the Rydberg atom, which scales with the principal quantum number $n$, as $n^2$ and can exceed one micrometer. This class of molecules has attracted significant attention because it demonstrates a new mechanism for chemical bonding and the molecules possess surprising features, such as the presence of a permanent electric dipole moment, even in the homonuclear case \cite{lpr11}. Here, we report the creation and theoretical description of ultralong-range Sr$_2$ molecules involving a $5sns$\,$^3S_1$ Rydberg atom.

Ultralong-range molecules were originally predicted theoretically \cite{gds00} and were subsequently observed in Rb \cite{bbn09} and Cs \cite{trb12}.  The original observations were of dimers involving spherically symmetric $S$ Rydberg states \cite{bbn09}, but now measurements have been extended to anisotropic $P$ \cite{bcb13} and  $D$ \cite{kgb14} Rydberg states and to molecules comprising one Rydberg atom and as many as five ground state atoms \cite{gkb14}.

There has been increasing interest in ultracold Rydberg gases of alkaline-earth metal atoms because of several new possibilities introduced by their divalent electronic structure. The principal transition of the Rydberg core is typically in the visible range and can be used to drive auto-ionizing transitions \cite{mlj10}, to image Rydberg atoms or ions \cite{mzs13}, and to provide oscillator strength for magic-wavelength optical trapping of Rydberg atoms \cite{mmn11}. Doubly excited states serve as strong perturbers of Rydberg states, creating a much richer assortment of electronic configurations than found in alkali-metal atoms. The existence of triplet and singlet excited levels provides many Rydberg series, giving access to attractive and repulsive interactions \cite{vjp12}. Two-photon excitation to triplet Rydberg levels via the intermediate $^3P_1$ state, as used here, can also reduce the overall decoherence from light scattering for a given strength of optical coupling to the Rydberg level as compared to two-photon transitions available in alkali-metal atoms \cite{hnp10}.  The results reported here represent the first experiments involving ultracold Rydberg atoms excited via intermediate triplet excited states.

Within the framework of a two-active-electron approximation, one of the two valence electrons can be excited to a Rydberg state.  The interaction between the excited Rydberg electron and a neighboring ground-state atom can be described using the Fermi pseudopotential \cite{fer34,omo77}

\begin{eqnarray} \label{Eq:Fermipseudo}
V_{pseudo}(\mathbf{r}_1,\mathbf{r}_2,\mathbf{R}) &=&  \sum_{i=1}^2 \frac{2\pi \hbar^2 A_s[k(\mathbf{R})]}{m_e}\delta(\mathbf{r}_i-\mathbf{R}) \\ \nonumber  &&+\frac{6\pi \hbar^2 A_p^3[k(\mathbf{R})]}{m_e}\overleftarrow{\nabla}\delta(\mathbf{r}_i-\mathbf{R})\overrightarrow{\nabla},
\end{eqnarray}
where $\mathbf{r}_i$ and $\mathbf{R}$ specify the positions of the Rydberg-atom valence electron and ground-state atom, respectively, measured from the Rydberg core. The momentum dependent $s$-wave and $p$-wave scattering lengths are $A_s(k)$ and $A_p(k)$. The Rydberg-electron momentum  in a semiclassical approximation is $\hbar k(\mathbf{r})=\sqrt{2m_e (e^2/(4\pi\varepsilon_0 \mathbf{r})-E_b)}$, where $E_b$ is the binding energy of the unperturbed Rydberg atom. This approximation is justified for highly excited $^3S_1$ Rydberg states because the electron-electron interaction is relatively small. For the case of both weak molecular binding compared to the binding energy of the Rydberg electron and a molecular size much larger than the characteristic size of the ground-state atom \cite{bbn09}, a first-order perturbative approximation of the interaction energy leads to the mean-field molecular potential

\begin{eqnarray}\label{Eq:MolPotential}
V(\mathbf{R})&=&\frac{2\pi \hbar^2 A_s[k(\mathbf{R})]}{m_e} \left[\int d\mathbf{r}_2 |\Psi(\mathbf{r}_1=\mathbf{R},\mathbf{r}_2)|^2\right. \\ && \left. + \int d\mathbf{r}_1 |\Psi(\mathbf{r}_1,\mathbf{r}_2=\mathbf{R})|^2\right], \nonumber
\end{eqnarray}
where $|\Psi(\mathbf{r}_1,\mathbf{r}_2)|^2$ is the valence-electron probability density for the Rydberg atom. The small correction due to $p$-wave scattering has been dropped in Eq.\ \ref{Eq:MolPotential}. For $A_s<0$, the interaction is attractive and can bind ground-state atoms to a Rydberg atom in potential wells created by the antinodes of the electron wavefunction. For Sr, the theoretically predicted value of the e$^-$-Sr scattering length is $A_s(k=0)=-18\,a_0$ \cite{bsa03}.

The creation of ultralong-range molecules requires ultracold temperatures so that thermal energies are lower than their small binding energies ($\sim 10$\,MHz). Also, high-density samples are necessary to ensure a sizeable probability of finding two atoms with separations less than the radial extent of a Rydberg electronic wavefunction.  We obtain these conditions using $^{84}$Sr atoms confined in an optical dipole trap (ODT).  This isotope has collisional properties favorable for evaporative cooling and the creation of high phase-space-density samples \cite{mmp08}. Details of the cooling and trapping are given in Ref.\,\cite{mmy09}.

At the start of the excitation time, the atoms are held in an ODT formed by crossed 1064-nm laser beams with waists of $300\,\mu\mathrm{m}\times 440\,\mu\mathrm{m}\times 38\,\mu\mathrm{m}$ resulting in an approximately oblate spheroidal trap with axial frequencies of 11\,Hz and 12\,Hz, and a radial frequency of 158\,Hz.  Typically $7\times 10^5$  atoms are trapped at a temperature of $200\,$nK yielding a peak density of $\rho=2.7\times10^{12}\,\mathrm{cm}^{-3}$. This corresponds to an average interparticle spacing of $1/\rho^{1/3}=0.7\,\mu$m, which is about $10\times$ the radius of the outer lobe of the Rydberg wavefunction $2n^{*2}a_0 \simeq 75$\,nm for $n=30$ and $n^*=n-\delta$.  We determine the $5sns$\,$^3S_1$ state quantum defect $\delta=3.372 \pm 0.001$ by fitting observed Rydberg lines between $n=24$ to 36, and this value agrees well with Ref. \cite{vjp12}.

Atoms are promoted to Rydberg states through two-photon $5s^2$\,$^1S_0$-$5s5p$\,$^3P_1$-$5sns$\,$^3S_1$ excitation. The $689$\,nm laser for the first step is detuned 170\,MHz to the blue of the intermediate state to avoid scattering from the atomic line and associated molecular resonances. Rydberg states with $n=29-36$ are reached with  photons at 319\,nm generated by frequency doubling the red output of a fiber-based optical parametric oscillator laser. Approximately 200\,mW of UV power is available. The intensities of the red and UV  light on the atoms are $2.2 \times10^3\,$W/m$^2$ and $2.3 \times10^5\,$W/m$^2$ respectively.  The UV and 689\,nm lasers co-propagate with orthogonal linear polarizations.  This configuration excites a superposition of $m = +1$ and $m=-1$ $^3S_1$ Rydberg states.

The frequency of the UV light is controlled by locking the 638\,nm fundamental to an optical cavity stabilized to the 689\,nm laser, which is locked to the $5s^2$\,$^1S_0$-$5s5p$\,$^3P_1$ atomic transition. The UV frequency is scanned using an acousto-optic modulator in a double-pass configuration in the path of the 638\,nm light en route to the stabilization cavity.   The excitation time is precisely controlled using an acousto-optic modulator on the 689\,nm beam, while the UV light is controlled by a slower mechanical shutter.  After excitation, the atoms are released from the trap and the ground-state atom population is measured with time-of-flight absorption imaging on the $5s^2$\,$^1S_0$-$5s5p$\,$^1P_1$ transition at 461\,nm. Excitation to atomic Rydberg or molecular levels is detected as ground-state atom-loss. The exposure time is adjusted for about 50\% peak loss, which is $\sim10$\,ms for atomic resonances and $\sim2$\,s for molecular resonances.  The ODT is left on during excitation, and we assume the AC Stark shift is the same for the atomic and molecular transitions in our quoted results.

\begin{figure*}[t]
\includegraphics[clip=true,keepaspectratio=true,width=5in,trim=0in 0in 0in 0in]{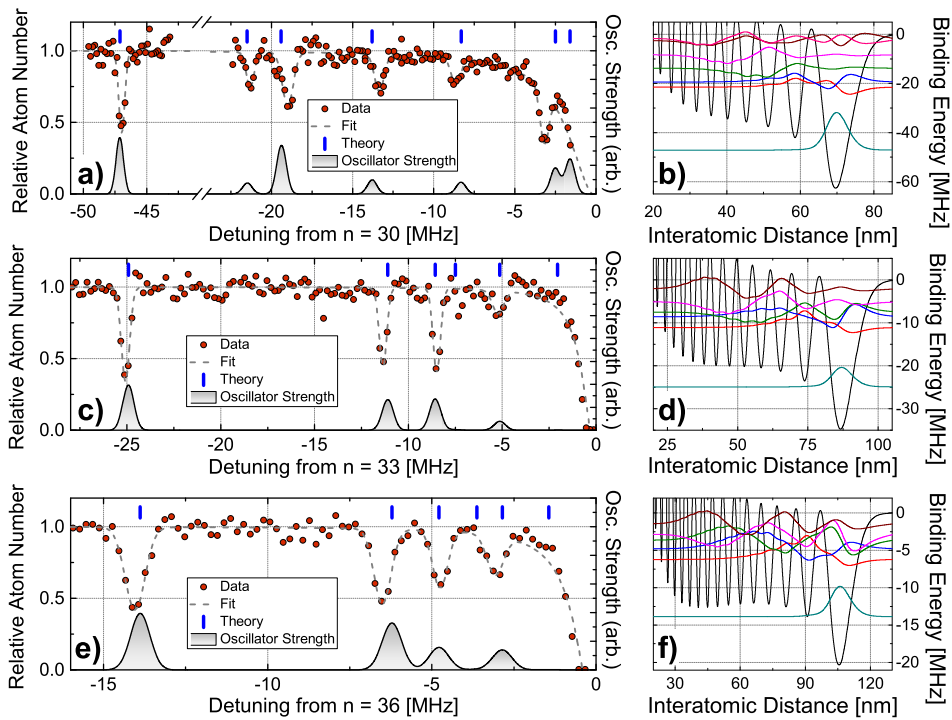}
\caption{(Left) Atom-loss spectra and (Right) calculated potentials and wavefunctions,$R \chi_\nu(R)$, for n=30 (top), 33 (middle), and 36 (bottom). The bars at the top of the molecular spectra indicate the positions of the theoretically-predicted binding energies of states bound by $> 1$\,MHz and the curves at the bottom indicate their calculated excitation strength. The origin of each frequency axis is  set to the center of the  atomic excitation spectrum (see text).
\label{Fig:Spectrum}}
\end{figure*}

Typical atom-loss spectra are shown in Fig.\ \ref{Fig:Spectrum}. The spectra are relatively simple because of the closed-shell $^1S_0$ electronic-ground state and lack of nuclear spin for $^{84}$Sr. Atom loss is an indirect method of detecting excitation as compared to the traditional technique of pulsed-field ionization and charged-particle detection \cite{bbn09,trb12,bcb13,kgb14,gkb14}, but it can still yield a high signal-to-noise ratio. A comprehensive understanding of the decay channels of ultralong-range molecules is still lacking, but it is clear that a large fraction of excitations should lead to measurable ground-state atom loss. Fluorescent decay after either atomic or molecular excitation has a sizable probability, approximately two-thirds, of creating long-lived $^3P_0$ or $^3P_2$ atoms.  Such atoms may remain trapped but they are invisible to absorption imaging.  The recoil energies of ${h}^2/2 m_{84} \lambda_{689}^2k_B=0.24\,\mu$K and ${h}^2/2 m_{84} \lambda_{320}^2k_B=1.1\,\mu$K are smaller than the $1.9\,\mu$K depth of the ODT.  However, collisional processes involving ground-state atoms and tunnelling to small internuclear separation can reduce the molecular lifetime significantly compared to the atomic lifetime for conditions similar to ours \cite{bbn11}. Such tunnelling should release enough energy to eject one or both atoms from the trap.  For atomic excitation, it is possible that the density of Rydberg atoms becomes high enough for inelastic Rydberg-Rydberg collisions to lead to atom loss \cite{rob05}. The density of molecules, however, is low enough that such processes are negligible.

The threshold for the molecular-binding energy for each principal quantum number is determined by measuring the resonance position for atomic excitation to the $5sns$\,$^3S_1$ state. This is done with a 10\,ms excitation time to avoid saturating the transition, and it results in a $\sim$ 800\,kHz FWHM linewidth, which is likely limited by the UV laser linewidth. For such short excitation times, no molecular transitions are visible.  For excitation times on the order of 1\,s, however, clearly resolved resonances corresponding to molecular bound states appear to the red of the (highly-saturated) atomic line. No transitions are observed to the blue or further to the red of the regions shown. Data are fit to the wing of a Lorentzian to describe the atomic background plus a gaussian for each molecular line.  Molecular-binding energies are determined by the frequency difference between molecular and atomic lines.

Typical molecular linewidths are 800 kHz FWHM, again limited by the UV laser linewidth. Several spectra were recorded for each principal quantum number, alternating between measurement of atomic and molecular lines. The uncertainties in the molecular line positions with respect to the atomic line, $\pm 150\,kHz$, are the statistical $1\sigma$ uncertainties in the mean for each group of measurements. DC Stark shifts of the transitions were measured for $n=36$ by applying  DC electric fields of up to about 0.5 V/cm. The extracted atomic and molecular DC polarizabilities were equal at our level of precision $\pm 0.5$\,MHz/(V/cm)$^2$ and calculated theoretically to be -4.5\,MHz/(V/cm)$^2$.

For the present range of $n$ and our relatively low densities, the production rate for trimers is expected to be much lower than for dimers  \cite{bbn10,gkb14} and therefore difficult to detect with our current methods.  The most deeply bound level observed  for each principal quantum number is assigned to the vibrational ground state  of one atom in the potential well formed by the outermost lobe of the Rydberg electron wavefunction. This is confirmed by calculations of the potentials and molecular wavefunctions (Fig.\ \ref{Fig:Spectrum}(Right)), which show that this state is well localized in this well. The more weakly bound levels correspond to vibrationally excited states, which are delocalized across several lobes of the electron probability density.

The electron wave function $\Psi(\mathbf{r}_1,\mathbf{r}_2)$ of the $^3S_1$ Rydberg atom can be calculated by numerically diagonalizing the Hamiltonian within the two-active-electron model. Spin-orbit interaction is included, but its effects are small \cite{vjp14}. The model potential is fitted to reproduce the measured energy levels in the singlet sector~\cite{ye14}, and yields the quantum defect $\delta = 3.376$ for $^3S_1$ states, which agrees well with the measured value.  The calculated wave functions for $^3S_1$ Rydberg states with $n \sim 30$ are dominated by a single configuration

\begin{equation}\label{Eq:3S1Wavefunction}
\Psi(\mathbf{r}_1,\mathbf{r}_2) \simeq \frac{1}{\sqrt{2}} \left(\phi_{5s}(\mathbf{r}_1) \psi_{ns}(\mathbf{r}_2)- \psi_{ns}(\mathbf{r}_1) \phi_{5s}(\mathbf{r}_2)\right)
\end{equation}
where $\phi_{5s}(\mathbf{r})$ and $\psi_{ns}(\mathbf{r})$ are the wave functions of the $5s$ state for Sr$^{+}$ ions and the Rydberg $ns$ state for Sr atoms, respectively. The contributions from the other configurations are smaller than 0.01\%. Considering that the wave function of the $5s$ state is rather localized (i.e. $\phi_{5s}(\mathbf{r}) \simeq 0$ for $r > 20$~a.u.), the molecular potential (Eq. \ref{Eq:MolPotential}) around $R \simeq 1000$~a.u. can be evaluated similarly to that for single-electron systems,

\begin{equation}\label{Eq:MolPotential2}
V(R) = \frac{2 \pi \hbar^2 A_s(k)}{m_e} |\psi_{ns}(\mathbf{R})|^2 + \frac{6 \pi \hbar^2 A_p^3(k)}{m_e} |\nabla \psi_{ns}(\mathbf{R})|^2 \, .
\end{equation}

By solving the Schr\"odinger equation associated with the molecular potential, the binding energies and the wave functions for Rydberg molecules are obtained.  The effective scattering lengths are taken to be

\begin{equation}\label{Eq:scatteringlength}
A_s(k) = A_{s}(k=0) + \frac{\pi}{3} \alpha k^2 \, \quad A_p(k) = A_{p}(k=0) \, .
\end{equation}

The measured value, $\alpha=186 a_0^3$, is used for the polarizability \cite{smb74}.  It is known \cite{bbn10} that a non-perturbative Green's function calculation correctly reproduces the measured molecular energy levels using the true zero-energy s-and p-wave scattering lengths. Within the first-order perturbative approximation, however, good agreement can be obtained by considering the scattering lengths as effective fitting parameters. For strontium, while the calculated $s$-wave scattering length is $A_{s}(k=0)=-18\,a_0$ \cite{bsa03}, an effective scattering length of $A_{s}(k=0) = -13.2 a_0$ for $n=30$ to $-13.3 a_0$ for $n=36$ yields good agreement with the measured energy levels, especially the most deeply bound molecular states. Although the depth of the deepest well in the molecular potential linearly scales with $A_{s}(k=0)$ (Eq.~\ref{Eq:MolPotential2}), the additional contribution from $p$-wave scattering becomes non-negligible around the nodes (i.e. $|\psi_{ns}({\bf R})| \simeq 0$) and affects the energies of the weakly bound levels and their density of states. They are optimally fitted using the value of $A_{p}(k=0) \simeq 8.4a_0$.

The molecular formation rate can be calculated as $\Gamma_\nu \propto |\langle \Psi, \chi_\nu | T | \Psi_0, \chi_0 \rangle|^2$ where $T$ is the transition matrix for two-photon absorption, and $\Psi_0$, $\chi_0$, $\chi_\nu$ are the ground-state-electron wavefunction, the pair distribution of ground-state atoms, and the $\nu$-th vibrational wavefunction of the Rydberg molecule, respectively. Within first-order perturbation theory, the electronic part of the transition matrix is independent of the molecular states. Therefore, the excitation rate is reduced to the Franck-Condon factor

\begin{equation}
\Gamma_\nu \propto \left|
  \int dR \, R^2 \, \chi_\nu(R) \chi_0(R)
\right|^2 \, .
\end{equation}
Moreover the pair distribution $\chi_0(R)$ is approximated as constant at large distances and can also be factored out.  The excitation rate $\Gamma_\nu$ is convolved with a gaussian distribution and the obtained excitation spectra are included in Fig.\ \ref{Fig:Spectrum}. The agreement between the measured and calculated spectra is very good.  The calculations reveal the existence of more molecular levels, but their excitation strengths are too weak to detect in the measurements. 

\begin{figure}[]
\includegraphics[clip=true,keepaspectratio=true,width=3in,trim=0in 0in 0in 0in]{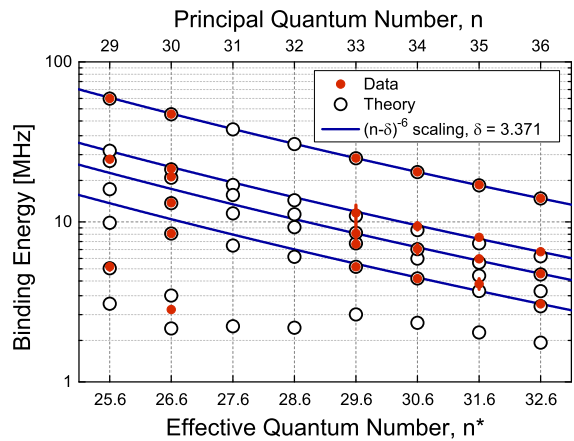}
\caption{Scaling of observed  molecular binding energies, showing $-1/(n-\delta)^6$ scaling for higher quantum numbers and more deeply bound levels.
 \label{Fig:Scaling} }
\end{figure}

Figure \ref{Fig:Scaling} shows the $n$ scaling of the observed binding energies together with the calculated values. The binding energies of deeply bound states follow the approximate $-1/(n-\delta)^6$ scaling  seen in Rb \cite{gkb14}.  Gaps in laser coverage prevented measurement of spectra for $n=31$ and $n=32$. Deviations from the scaling are evident for weakly bound states and are clearly seen in high-quality spectra for $n=30$. The energy scaling reflects  the scaling of the probability density $|\psi_{ns}(\mathbf{R})|^2$ of the Rydberg electron at the location of the unperturbed Sr atom. The wave functions of the most deeply bound states for $29 \le n \le 36$ are confined in the most outer well of the molecular potential (Fig.\ \ref{Fig:Spectrum}). Therefore, the corresponding energies scale with the depth of the well by $-1/(n-\delta)^6$ as the quantum number $n$ varies. When the first-excited state of this potential well is nearly degenerate with the lowest energy state of the adjacent well, the coupling between two states yields an energy splitting, $\Delta$. If a molecular state confined in a single well for a given $n$ experiences tunneling to an adjacent well for another value of $n$, the molecular energy does not scale with the potential depth showing a deviation $\sim \Delta$ from the $-1/(n-\delta)^6$ scaling. Such a transition in the wave function is seen for some excited vibrational levels (Fig.\ \ref{Fig:Spectrum}). For example, although the first excited vibrational state's wave function is rather confined within the first two outer wells for the given range of $n$, the second-excited state tunnels ever deeper into inner wells as $n$ increases. Thus the deviation from the scaling is prominent for the latter (Fig.\ \ref{Fig:Spectrum}). Note that data obtained for $n=29$ were of inferior quality, but we were able to identify the three lines indicated in Fig.\ \ref{Fig:Scaling}.

We have presented the first observation of ultralong-range Rydberg molecules in Sr$_2$ formed by photoexcitation to the $5sns$\,$^3S_1$ Rydberg state for $29 \leq n \leq 36$. The observed lines are well described using a Fermi pseudopotential approach to calculate perturbative molecular potentials and yield the effective $s$-wave and $p$-wave e$^-$-Sr scattering lengths $A_s=-13.2 a_0$ and $A_p=8.4a_0$.

This work represents the first study of ultralong-range Rydberg molecules in a divalent-atomic system, and it opens new directions in this emerging research area. Doubly excited electronic states give rise to dramatic perturbations of the Rydberg states in divalent atoms.  This should lead to new types of ultralong-range molecules with mixed electronic character that arise from degeneracies of pairs of Rydberg levels of different angular momenta. This may also lead to very strong transition strengths for production of molecules with large dipole moments. If one can form Rydberg molecules with high electronic angular momenta, it might be possible to optically trap them using the oscillator strength of the ionic core \cite{mmn11}. High densities of atoms in metastable triplet levels can be created in these systems \cite{tcm08}, allowing the formation of Rydberg molecules in which triplet atoms serve as the ``ground-state" atoms. Spectroscopy of these molecules will probe the low-energy scattering of electrons from the metastable states including measuring the electron-triplet scattering length, which should be sensitive to the greater polarizability compared to closed-shell, ground-state atoms.

Detection of ultralong-range molecules with atom loss, as demonstrated here, greatly simplifies the required experimental apparatus compared to charged particle detection, which may open the way for study of these exotic molecules in many other species of atomic gases. The measurements reported here also represent an important step towards future experiments with interacting alkaline-earth Rydberg atoms because molecular excitations represent loss channels that need to be avoided.

\vspace{0 in}

\textmd{\textbf{Acknowledgements}}

Research supported by the AFOSR under grant no. FA9550-12-1-0267, the NSF under grants nos. 1301773 and 1205946, and the Robert A. Welch Foundation under grants nos. C-0734 and C-1844, and by the FWF (Austria) under grant no P23359-N16 and by the SFB-NextLite.  The Vienna Scientific Cluster was used for the calculations.

\end{document}